\documentclass{article}
\usepackage{graphics}
\usepackage{graphicx}
\usepackage{amssymb}
\usepackage{amsmath}
\usepackage{amsfonts}
\usepackage{parskip}
\usepackage{cite}
\usepackage{epsfig}
\usepackage{float}
\usepackage{fontenc}
\usepackage[latin1]{inputenc}
\usepackage{times}
\usepackage[english]{babel}

\begin{document}
\title{Non-isothermal transport of multi-phase fluids in porous media. Constitutive equations}
\author{Signe Kjelstrup$^a$, Dick Bedeaux$^a$, Alex Hansen$^b$, Bj{\o}rn
Hafskjold$^a$, Olav Galteland$^a$ \\
PoreLab, $^a$Department of Chemistry, $^b$Department of Physics \\
Norwegian University of Science and Technology, NTNU, 7491 Trondheim}
\maketitle
\date{}

\maketitle

\begin{abstract}

{We develop constitutive equations for multi-component, multi-phase, 
macro-scale flow in a porous medium exposed to temperature-, composition-, and pressure -gradients. The porous medium is non-deformable. We define the pressure and the composition of the representative elementary volume (REV) in terms of the volume and surface averaged pressure and the saturation, and the respective driving forces from these variables. New contributions due to varying porosity or surface tension offer explanations for non-Darcy behavior. The interaction of a thermal and mechanical driving forces give thermal osmosis. An experimental program is suggested to verify Onsager symmetry in the transport coefficients.}

\subsection*{Keywords:} porous media, pore-scale, energy dissipation, two-phase
flow, representative elementary volume, macro-scale, effective driving
forces, excess surface- and line-energies, pressure, non-equilibrium
thermodynamics 
\end{abstract}

\section{Introduction}

We have recently \cite{Kjelstrup2018} derived the entropy production, 
$\sigma $, of a representative elementary volume (REV) in a heterogeneous
system, a multi-phase, multi-component fluid in a porous non-deformable matrix. 
The coarse-grained description of the REV was formulated for systems
that obey Euler homogeneity. A Gibbs equation could therefore be formulated for the
REV itself, and used as a starting point, as is common in non-equilibrium thermodynamics \cite{deGroot1984}. Once the entropy production was 
found from this and the balance equations, the driving forces and the constitutive equations can be found. This will be done here. 

We shall see that the form of the equations is the same as for homogeneous systems, but the driving forces are particular for the multi-component, non-isothermal and
multi-phase flow in porous media. We are seeking internal relations between
experiments particular for these flows, as derived for instance from the
Onsager relations.

The procedure that we used to obtain the Gibbs equation for coarse-grained variables 
\cite{Kjelstrup2018} assumed that the additive thermodynamic variables of
the REV are Euler homogeneous functions of the first order. As it is so central to this work, we will briefly review the procedure in Section 2, highlighting the main points.  In the procedure, we regard the REV as a complete thermodynamic system. Hansen and Ramstad \cite{Hansen2009} suggested this  
possibility already some time ago. Since then the hypothesis has been
supported through measurements on Hele-Shaw cells \cite{Erpelding2013} and
through network simulations \cite{Savani2017}. The variables of the REV will
then fluctuate similar to the variables in a normal thermodynamic state around a mean value.

The equations presented here, will allow us to revisit previously published experimental results, and explain in more detail for instance when we can expect deviations from Darcy's law. It appears that the strictly linear theory
may cease to hold for small pressure differences; also for single fluids 
\cite{Swartzendruber1962,Miller1963,Boersma1973, Bernadiner1994}. 
Observations of deviations from Darcy's law were made for water or water solutions in clay \cite{Bernadiner1994,Boersma1973}. Thresholds and/or deviations from straight lines in plots of flow versus the overall pressure difference, were reported. 
Boersma et al. \cite{Boersma1973} found a dependency of
the threshold on the average pore radius, $\bar{r}$, for flow in a porous medium made of glass-beads. These observations have, as of yet, no unique explanation. 
When dealing with immiscible fluids, Tallakstad et al. \cite{Tallakstad2009} observed a square dependence of the flow rate on the pressure difference under steady-state flow conditions.  Sinha and Hansen 
\cite{Sinha2012} explained this square dependence by the successive opening of pores due to the mobilization of interfaces when the pressure difference across the sample is increased.  This explanation was supported by a mean-field calculation and 
numerical experiments using a network model.  Sinha et al. \cite{Sinha2012,Sinha2017} followed up the original Tallakstad study, originally done in a two-dimensional model porous medium, both experimentally and computationally in three-dimensional porous media,
with the same result.

There is not only a need to better understand deviations from Darcy's law
for volume transport. Other driving forces than those related to pressure
differences are also relevant to porous media. Counter-current transport of
components can lead to gradients in composition (chemical potential), or
chemical driving forces. Constant injection of cold seawater into a warm
hydrocarbon reservoir can create thermal driving forces. The presence of a
porous matrix has an impact on the flow pattern and for instance the Soret
coefficient \cite{Davarzani2010}. An emerging concept for water cleaning is based on thermal osmosis \cite{Keulen2017}. This process could help produce water using industrial and natural heat sources, a very important topic in the world today. 

It is still an open question in porous media theory, how driving forces, like 
the ones mentioned, interact in porous media, and how the porous medium make
these interactions special \cite{Davarzani2010}. It is the aim of this work to clarify the coupling that can take place due to these forces, by constructing a non-equilibrium thermodynamic theory, particular for porous media.  

The paper is structured as follows. Section \ref{subsec:ex} gives a  
brief repetition of the variables used to derive the entropy production, which follows in Section \ref{sec:ep}  \cite{Kjelstrup2018}. As before, we have assumed that the system obeys Euler homogeneity of the first order, meaning that we restrict ourselves to non-deformable media and a constant ratio of fluid surface area to volume (no film formation). For such systems we proceed to find detailed expressions for the part of the chemical potential of the REV. The driving forces, due to temperature -, pressure - and chemical
potential gradients, are specified. The simple case of flow of two immiscible fluids is used to bring out the specifics of porous media.  An experimental program is suggested in the end to verify Onsager symmetry in the transport coefficients.

\section{Thermodynamic variables for the REV}
\label{subsec:ex}

\begin{figure}[h]
\centering
\includegraphics[scale = 0.35]{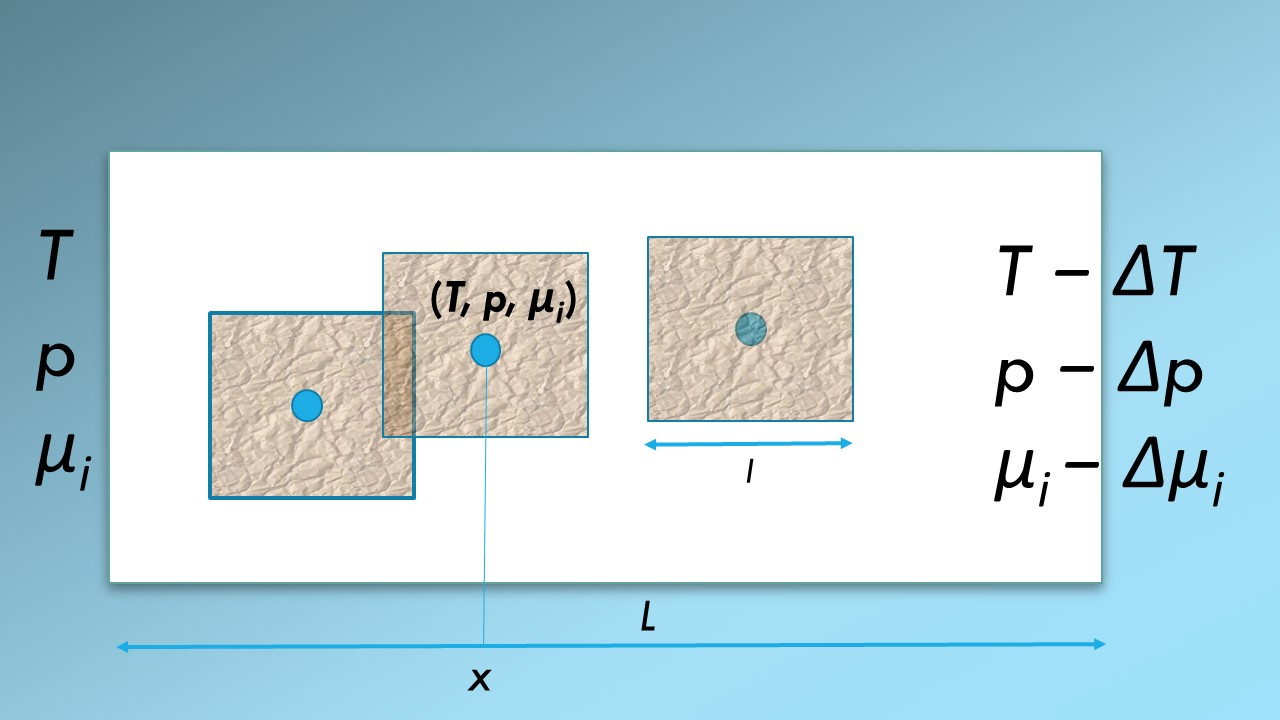}
\caption{A representative elementary volume (REV)(magenta squares, length $l$%
) in a system (white box of length $L$) that is exposed to for instance a
pressure difference, $\Delta p$, a temperature difference, $\Delta T$, and
differences in chemical potentials $\Delta \protect\mu _{i}$. A (blue) dot
is used to represent the state that characterizes the REV. The set of state
variables are positioned on the $x$-axis: The temperature, pressure and the
chemical potentials, {($T,p$, $\protect\mu _{i}$})(x).}
\label{fig:REV2}
\end{figure}

The central concept in this analysis is the representative elementary
volume; the REV \cite{hg1990,gh1998}. Its characteristic size, $l$, is small
compared to the size (length) of the full system, $L$, but large compared to
the characteristic pore length and diameter. An illustration of the REV is
given by the squares in Fig. \ref{fig:REV2}. The REV (square) consists of
several phases and components. The problem is to find the representation on
the larger scale. For each point in the porous system, represented by the
(blue) dot in Fig.\ref{fig:REV2}, we use the REV around it to obtain the
variables ($U^{\text{REV}},S^{\text{REV}},V^{\text{REV}},M_{i}^{\text{REV}}$%
) of the REV. This was first defined in \cite{Kjelstrup2018}. From the Euler
homogeneity of these variables, the possibility followed to define also
the temperature, pressure and chemical potentials of the REV, ($T,p,\mu _{i}$) 
\cite{Kjelstrup2018}. 

The procedure will be recapitulated. We follow standard thermodynamics and
choose as a basis set of variables, the set that consists of internal
energy, entropy, volume, ($U,S,V$), and the component masses, $M_{i}$. These
variables, are given superscript REV, and constitute the only \textit{%
independent} variables of the REV.

The value of each of these variables of the REV is obtained as a sum of
contributions from each phase, interface and three-phase contact line
present \cite{hg1990,gh1998}. The contributions arise say from the phase
volumes, the surface areas and the contact line lenghts and are pore-scale
variables; they are not \emph{independent} variables on the macro-scale.
Assuming Euler homogeneity, means that a REV of the double size, for
example, has double the energy, entropy, and mass, as well as double the
surface areas of various types and double the line lengths. The average
surface area, pore length and pore radius, as well as the curvature of
the surfaces in the REV, are then everywhere the same.

A system of $k$ components in $m$ homogeneous phases, has a volume, $V^{%
\text{REV}}$, with contributions from the homogeneous bulk phases $V^{\alpha
,\text{REV}}$, $m\geq \alpha \geq 1$, and the excess line volumes, $%
V^{\alpha \beta \delta ,\text{REV}}$, $m\geq \alpha >\beta >\delta \geq 1$.%
\begin{equation}
V^{\text{REV}}=\sum_{\alpha =1}^{m}V^{\alpha ,\text{REV}}+\sum_{\alpha
>\beta >\delta =1}^{m}V^{\alpha ,\beta ,\delta ,\text{REV}}\approx
\sum_{\alpha =1}^{m}V^{\alpha ,\text{REV}}
\end{equation}%
The excess surface volumes are by construction zero. The excess line volumes
are not, because the dividing surfaces in general cross each other along
three different lines. We neglect these contributions, which are normally small also in porous media. The volume
of the pores is%
\begin{equation}
V^{p,\text{REV}}=\sum_{\alpha =1}^{m-1}V^{\alpha ,\text{REV}}
\end{equation}%
Superscript $p$ is used for pore. The porosity, $\phi $, and the degree
of saturation, $\hat{S}^{\alpha }$ (saturation for short), are 
\begin{equation}
\phi \equiv \frac{V^{p,\text{REV}}}{V^{\text{REV}}}\ \ \ \ \ \mathrm{{and}\
\ }\hat{S}^{\alpha }\equiv \frac{V^{\alpha ,\text{REV}}}{V^{p,\text{REV}}}=%
\frac{V^{\alpha ,\text{REV}}}{\phi V^{\text{REV}}}  \label{I.3}
\end{equation}%
Superscript $\alpha $ is used for a component, which is equal to the phase
in the present case. The porosity and the saturation do not depend on the
size of the REV, and have therefore no REV superscript. 

The mass of component $i$ in the REV, $M_{i}^{\text{REV}}$, is the sum of
the masses in the homogeneous phases of the REV, $\alpha $, $M_{i}^{\alpha ,%
\text{REV}}$, $m\geq \alpha \geq 1$, the excess interfacial masses, $%
M_{i}^{\alpha \beta ,\text{REV}}$, $m\geq \alpha >\beta \geq 1$, and the
excess line masses, $M_{i}^{\alpha \beta \delta ,\text{REV}}$, $m\geq \alpha
>\beta >\delta \geq 1$. We obtain: 
\begin{equation}
M_{i}^{\text{REV}}=\sum_{\alpha =1}^{m}M_{i}^{\alpha ,\text{REV}%
}+\sum_{\alpha >\beta =1}^{m}M_{i}^{\alpha \beta ,\text{REV}}+\sum_{\alpha
>\beta >\delta =1}^{m}M_{i}^{\alpha \beta \delta ,\text{REV}}  \label{V.1}
\end{equation}%

We shall often use the example of two immiscible one-component phases $w$
and $n$ in a solid porous material $r$ of porosity $\phi $, where contact
line contributions are negligible. We can think of phase $w$ as wetting, and $n$ as non-wetting. The mass variables are then from Eq. \ref%
{V.1}: 
\begin{equation}
M_{n}^{\text{REV}}=M_{n}\ \ ,\ \ M_{w}^{\text{REV}}=M_{w}+M_{w}^{wn}\ \ 
\text{and}\ \ M_{r}^{\text{REV}}=M_{r}+M_{r}^{rn}+M_{r}^{rw}
\end{equation}%
When an interface is formed between two phases, we are free to choose the
position of the interface such that one of the components has a zero excess
mass. This is the position of the equimolar surface of this component. This
position is convenient because the number of variables are reduced. When we
use the equimolar surface of $n$, $M_{n}^{rn}=M_{n}^{wn}=0$, and when we use the
equimolar surface of $w$ at the surface of the solid, $M_{w}^{rw}=0$.
These choices simplify the description of the REV. 

The expressions for $U^{\text{REV}}$ and $S^{\text{REV}}$ are similar to Eq. %
\ref{V.1}. This way to construct a REV is reminiscent of the geometric construction of a state function, proposed for flow in porous media by 
McClure \emph{et al.} \cite{McClure2018}.

The basis set of macro-scale variables of the REV ($U^{\text{REV}},S^{\text{%
REV}},V^{\text{REV}},M_{i}^{\text{REV}}$) apply to the whole REV. The
temperature, pressure and chemical potentials of the REV, ($T,p,\mu _{i}$),
on the macro-scale were next defined, as is normal in thermodynamics, as partial
derivatives of the internal energy. These definitions are normal in the
sense that they have the same form as they have in homogeneous systems. They differ
from definitions in normal homogeneous systems in that the variables here
(say $U^{\text{REV}}$) have contributions from all parts of the
heterogeneous REV. The intensive variables $T,p$ and $\mu _{i}$ are \emph{not}
averages of the corresponding variables on the pore-scale, as was also
pointed out by Hassanizadeh and Gray [11,12]. 

The macro-scale densities of internal energy, entropy and mass; in the
example, $u,s,\rho _{i}$, do not depend on the size of the REV. The
densities are therefore convenient when we need to integrate across the
system [1]. They are, however, functions of the position of the REV.

\section{The entropy production of two-phase flow}
\label{sec:ep}

Pressure-driven mass flow through porous media can lead to gradients in
composition and temperature, and vice versa; temperature gradients can lead
to mass flow, separation and pressure gradients. The interaction of such
flows is of interest, and motivated the search for convenient forms of
the entropy production \cite{Kjelstrup2018}. 

\subsection{Component flow variables}

From the Gibbs equation for the REV, we derived the entropy production for
transport of heat and two immiscible fluid phases through the REV 
\cite{Kjelstrup2018}. With transport in the $x$-direction only, the entropy
production $\sigma $ of the example system was 
\begin{equation}
\sigma =J_{q}^{\prime }\frac{\partial }{\partial x}(\frac{1}{T})-\frac{1}{T}%
\left( J_{w}\frac{\partial \mu _{w,T}}{\partial x}+J_{n}\frac{\partial \mu
_{n,T}}{\partial x}\right)  \label{E.7c}
\end{equation}%
The frame of reference for the mass transport is the non-deformable solid
matrix, $J_{r}\equiv 0.$ Here $J_{q}^{\prime }$ is the sensible heat flux in J/(m$^2$s)), $T $ is the temperature (in K), $J_{i}$ is a component flux (in kg.m$^{-2}$.s$^{-1}$) and $\partial \mu _{i,T}/\partial x$ is the gradient of chemical potential
(in J.kg$^{-1}$.m$^{-1}$) evaluated at constant temperature. All properties
are for the REV so the superscript is omitted. 

The thermal force conjugate to the heat flux is the gradient of the inverse temperature, where the temperature was defined for the REV as a whole, see \cite{Kjelstrup2018}. This force will not be further discussed. 

The chemical force conjugate to the mass flux is the inverse temperature times the
negative gradient of $\mu _{i,T}$. This driving force has several contributions, which must be defined for the REV. We are seeking more specific expressions for the last driving forces in Eq.\ref{E.7c}. 

In order to take all REV variables into account, see Section \ref{sec:G} below, we
derive the chemical potential starting from the Gibbs energy, $G$: 
\begin{equation}
G\equiv U+pV-ST=\sum_{i=1}^{k}\mu _{i}M_{i}=\sum_{i=1}^{k}G_{i}  \label{eq:G}
\end{equation}%
where $G$ applies to the REV and $G_{i}$ is defined for component $i$ in the last identity. The total differential of $U$ is introduced and we obtain 
\begin{equation}
dG=-SdT+Vdp+\sum_{i}\mu _{i}dM_{i}  \label{eq:dG}
\end{equation}%
The full chemical potential is the derivative with respect to $M_{i}$, which was defined in the previous section; 
\begin{equation}
\mu _{i}\equiv \left( \frac{\partial G}{\partial M_{i}}\right) _{T,p}  \label{eq:dmu}
\end{equation}%
The total differential of the chemical potential is: 
\begin{equation}
d\mu _{i}=-S_{i}dT+V_{i}dp+\sum_{j=1}^{k}\mu _{i,j}^{c}dM_{j}\equiv
-S_{i}dT+V_{i}dp+d\mu _{i}^{c}  \label{eq:mui}
\end{equation}%
where $S_{i}=-\left( {\partial \mu _{i}}/{\partial T}\right) _{p,M_{j}}$ and 
$V_{i}=\left( {\partial \mu _{i}}/{\partial p}\right) _{T,M_{j}}$ and $\mu
_{i,j}^{c}=({\partial \mu _{i}}/{\partial M_{j}})_{p,T,M_{k}}$ are partial
specific quantities. The last term describes the change in the chemical
potential by changing composition of the medium. The $d\mu _{i,T}$ is now
defined as a part of the whole differential: 
\begin{equation}
d\mu _{i,T}\equiv d\mu _{i}+S_{i}dT=V_{i}dp+d\mu _{i}^{c}  \label{eq:muiT}
\end{equation}%
The last term is zero when the composition is uniform. The pressure gradient
can be introduced as a driving force in Eq. \ref{E.7c} through this equation.

\subsection{Volume flow as variable}

The volume flow is often measured, and is thus a central variable. 
In the simple case of a single fluid, say w, we obtain from Eq.\ref{E.7c} and \ref{eq:muiT} that
\begin{equation}
\sigma = J_{q}^{\prime }\frac{\partial }{\partial x}(\frac{1}{T})- J_V\frac{1}{T}%
\frac{\partial p}{\partial x} \label{E.7d}
\end{equation}%
The volume flow is $J_V=J_wV_w$ for a single fluid. For two fluids 
\begin{equation}
J_{V}\equiv J_{n}V_{n} + J_{w}V_{w}
\end{equation}
In the absence of a gradient in composition, $d\mu_i^c=0$, the same expression applies with this volume flow. 
We may also follow Hansen et al \cite{Hansen2018} and write the component contributions as 
$J_{n}V_{n}=\hat{S}_{n}v_{n}$, $J_{w}V_{w}=\hat{S}
_{w}v_{w}$ and $J_{V}=v=\hat{S}_{n}v_{n}+\hat{S}_{w}v_{w}$.
where the saturation has been introduced, and $v_i$ is the volume flow of i. 

With two fluids in a uniform, non-deformable rock, there are three components.
On the coarse-grained level, these are mixed. We assume that $d\mu
_{r}^{c}=0$, and obtain Gibbs-Duhem's equation on the form  
\begin{equation}
\rho _{n}d\mu _{n}^{c}+\rho _{w}d\mu _{w}^{c}=0  \label{GD}
\end{equation}%
where $\rho _{i}$ is the density of $i$ in the REV (in kg.m$^{3}$). 

This can be used with Eq.\ref{eq:muiT} and $J_{V}$ to change Eq.\ref{E.7c} into 
\begin{equation}
\sigma =J_{q}^{\prime }\frac{\partial }{\partial x}\left( \frac{1}{T}\right)
-J_{V}\frac{1}{T}\frac{\partial }{\partial x}p-J_{D}\frac{\rho _{w}}{T}\frac{%
\partial \mu _{w}^{c}}{\partial x}  \label{D.1}
\end{equation}
The entropy production is invariant, and this defines $J_{D}$ as the velocity of component $w$ relative to $n$ 
(in m.s$^{-1}$): 
\begin{equation}
J_{D}\equiv \frac{J_{w}}{\rho _{w}}-\frac{J_{n}}{\rho _{n}}
\end{equation}%
The entropy production \ref{D.1} has also three terms. While the first term
on the right-hand side is the same as before, the second term is the volume
flow with minus the pressure gradient over the temperature as driving force,
and the third term is the velocity difference with the chemical potential
gradient times the density over the temperature as driving force.

Equations \ref{E.7c} and \ref{D.1} are equivalent. They describe the same
entropy production or flow dissipation. They provide alternative choices of
conjugate thermodynamic force-flux pairs. The choice to use in the
particular case, is determined by practical reasons; what can be measured or
not, which terms are zero. For instance, under isothermal conditions we need
not take the term containing the heat flux along, even if heat may be
transported reversibly. One set may give a negative contribution to the entropy production (work is done), but the overall entropy production is positive, of course. Each set can be used to obtain constitutive equations for transport on the macro-scale. We shall proceed to find these for porous media flow, finding first more detailed expressions for the driving forces. In order to do so, we again use the additive properties of the REV-variables.

\section{The chemical driving force}
\label{sec:G}

\subsection{Saturation-dependent contributions}

We need the specific contribution to the chemical potential gradients in the
entropy production in Eqs.\ref{E.7c}. In order to see the meaning of $\mu
_{i}$ in a porous medium, we study the contributions to the Gibbs energy in
more detail. The component contributions in the REV are additive, cf. Eq. \ref%
{V.1}. In principle, each component can exist in all phases in the REV. For
component $i$ we therefore have 
\begin{eqnarray}
G_{i}^{\mathrm{REV}} &\equiv &\mu _{i}^{\mathrm{REV}}M_{i}^{\mathrm{REV}} 
\notag \\
&=&\sum_{\alpha =1}^{m}G_{i}^{\alpha ,\text{REV}}
+\sum_{\alpha >\beta=1}^{m}G_{i}^{\alpha \beta ,\text{REV}}  \notag \\
&=&\sum_{\alpha =1}^{m}g_{i}^{\alpha }V^{\alpha ,\text{REV}}
+\sum_{\alpha>\beta =1}^{m}g_{i}^{\alpha \beta }\Omega ^{\alpha \beta,\text{REV}}
\end{eqnarray}%
The expression gives the Gibbs energy contributions of component $i$ to the
REV. We neglected again possible contributions from contact lines.

In the case of two immiscible, one-component fluids in a non-deformable
porous rock, we obtain for the non-wetting fluid. 
\begin{eqnarray}
G_{n}^{\mathrm{REV}} &\equiv &\mu _{n}M_{n}^{\mathrm{REV}}=G_{n}^{n,\text{REV%
}}+G_{n}^{rn,\text{REV}}+G_{n}^{wn,\text{REV}}  \notag \\
&=&g_{n}^{n}V^{n,\mathrm{REV}}+g_{n}^{rn}\Omega ^{rn,\text{REV}%
}+g_{n}^{wn}\Omega ^{wn,\text{REV}}  \label{M.2}
\end{eqnarray}%
For immiscible, one-component fluids the label indicating the components
also gives the phase. The densities $g_{n}^{n},\ g_{n}^{rn}$ and $g_{n}^{wn}$
are averages over $V^{n,\mathrm{REV}},\ \Omega ^{rn,\text{REV}}$ and $\Omega
^{wn,\text{REV}}$. Their local values in the pores may vary around these
averages.

The concentration dependent part of the chemical potential of $i$, $\mu _{i}$%
, is (in J.kg$^{-1}$) 
\begin{equation}
\mu _{i}=\mu _{i}^{0}+\frac{RT}{W_{i}}\ln \frac{\rho _{i}}{\rho _{i}^{0,%
\text{REV}}}  \label{17}
\end{equation}%
Here $R$ is the gas constant (in J.K$^{-1}$.mol$^{-1}$) and $W_{i}$ is the
molar mass (in kg.mol$^{-1}$). The chemical potential is measured referred
to a standard state, $\mu _{i}^{0}$, having the local concentration $\rho
_{i}^{0}$ in all the pores. In the description of porous media, a convenient reference state may be the state when one component is filling all pores, or 
the saturation is unity, $\hat{S}^{i}=1$. The
mass density of $i$ in the REV for the standard state is $\rho _{i}^{0,\text{%
REV}}=\rho _{i}^{0}V_{p}^{\text{REV}}/V^{\text{REV}}=\rho _{i}^{0}\phi $. 
Away from this state $\rho _{i}=\rho _{i}^{0}\hat{S}^{i}\phi$ 
for $i=n,w$. This gives
\begin{equation}
\frac{\rho_i}{\hat{S}^i} = \rho^0_i \phi
\end{equation}
By introducing these definitions into Eq.\ref{17}, we obtain for the
concentration dependent part of the driving force 
\begin{equation}
\frac{\partial \mu _{i}^{c}}{\partial x}=\frac{RT}{W_{i}}\frac{1}{\hat{S}^{i}%
}\frac{\partial \hat{S}^{i}}{\partial x}  \label{mu}
\end{equation}%
We see that any variation in saturation between REVs along the system, will
lead to a driving force. We integrate between two REVs,  and we obtain the chemical driving force for porous media flow
\begin{equation}
\rho_i \Delta \mu _{i}^{c} = \phi \frac{\rho^0_iRT}{W_i} \Delta \hat{S}_i
\label{eq:mu_c}
\end{equation}
We have seen that the return to the Gibbs energy of the porous medium helped define the chemical potential in terms of properties relevant to porous media. All variables are measurable.  

\subsection{The pressure of a REV}
\label{REVpressure}

We find the contribution from the pressure to the driving force, by starting as above with the extensive property that holds the variable. For the pressure, this is the grand potential. The compressional energy of the REV is equal to minus the grand potential: 
\begin{equation}
\Upsilon ^{\text{REV}}\left( T,V^{\text{REV}},\mu _{i}\right) \equiv -pV^{%
\text{REV}}=U^{\text{REV}}-S^{\text{REV}}T-\sum_{i=1}^{k}\mu _{i}M_{i}^{%
\text{REV}}
\end{equation}%
The grand potential of the REV is additive, which gives 
\begin{equation}
\Upsilon ^{\text{REV}}=\sum_{\alpha =1}^{m}\Upsilon ^{\alpha ,\text{REV}%
}+\sum_{\alpha >\beta =1}^{m}\Upsilon ^{\alpha \beta ,\text{REV}%
}+\sum_{\alpha >\beta >\delta =1}^{m}\Upsilon ^{\alpha \beta \delta ,\text{%
REV}}
\end{equation}%
We introduce contributions from all phases, surfaces and lines. This gives
for the pressure of the REV: 
\begin{equation}
p=\frac{1}{V^{\text{REV}}}\left( \sum_{\alpha =1}^{m}p^{\alpha }V^{\alpha ,%
\text{REV}}-\sum_{\alpha >\beta =1}^{m}\gamma ^{\alpha \beta }\Omega
^{\alpha \beta ,\text{REV}}-\sum_{\alpha >\beta >\delta =1}^{m}\gamma
^{\alpha \beta \delta }\Lambda ^{\alpha \beta \delta ,\text{REV}}\right) 
\label{eq:PVREV}
\end{equation}%
The superscript denotes the relevant phase. The last equation makes it
possible to compute the pressure of the REV, $p$, from the pressures in the
bulk phases, and the surface- and line tensions. With knowledge of the
pressure in the REV, we find the driving force, $-dp/dx$, in the entropy
production, Eq.\ref{D.1}. We explain now how we can define and compute the
pressure from Eq.\ref{eq:PVREV}, using the example of two immiscible fluid
in a non-deformable medium.

\begin{figure}[t]
\centering
\includegraphics[scale = 0.35]{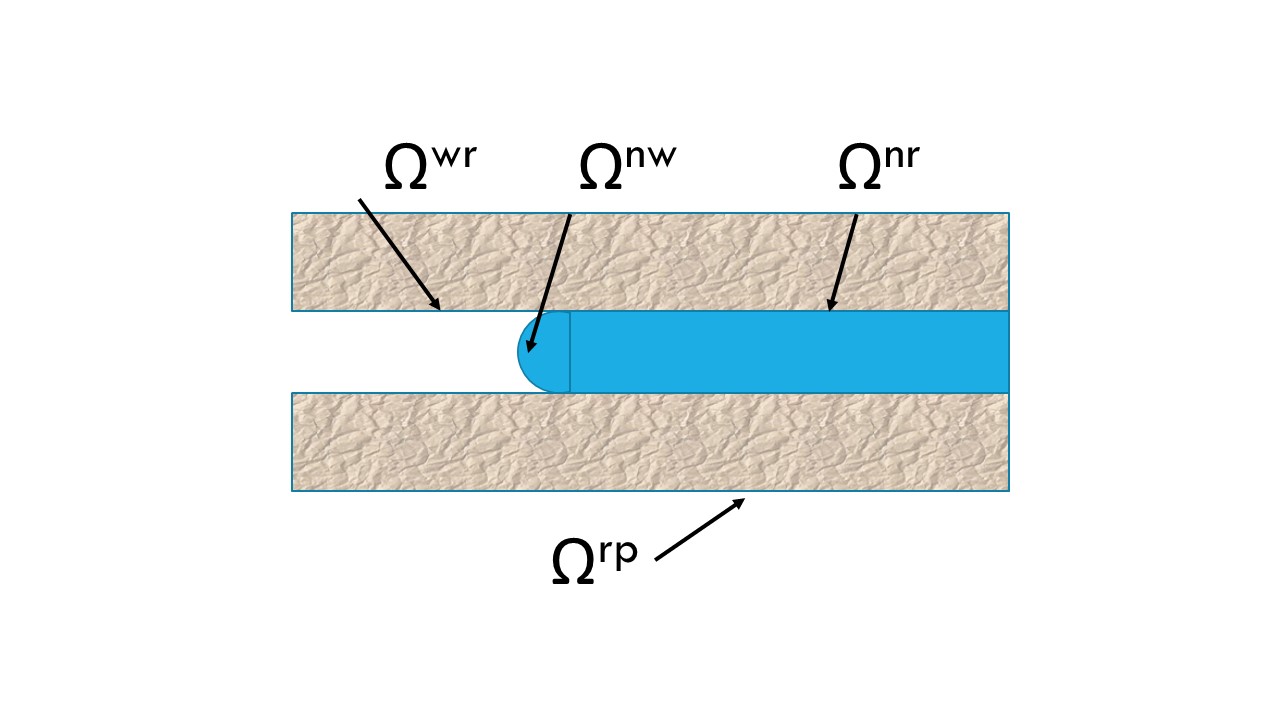}
\caption{Illustration of contact areas between the phases. A cylindrical
pore is chosen for simplicity}
\label{fig:pore-surface-areas}
\end{figure}

We follow Eq.\ref{eq:PVREV} and sum over the $n$, $w$ and $r$ bulk phases,
and the $nr,wr$ and $nw-$interfaces. The situation can be illustrated for a
single cylindrical pore, see Fig. \ref{fig:pore-surface-areas}. 

The figure shows two phases $n$ and $w$ in a tube with the average radius.
The wall material is $r$. Contact areas are therefore $\Omega ^{nr,\text{REV}%
}$, $\Omega ^{wr,\text{REV}}$, and $\Omega ^{nw,\text{REV}}$. The total
surface area of the pore is $\Omega ^{rp,\text{REV}}\equiv \Omega ^{nr,\text{%
REV}}+\Omega ^{wr,\text{REV}}$. The area $\Omega ^{nw,\text{REV}}$ is the
smallest contact area shown in the figure. The volumes in Eq. \ref{eq:PVREV}
depend on the saturation of the non-wetting component, $\hat{S}^{n}$ and the
porosity, $\phi $. Neither of the fluids form a film between the surface and
the other fluid, so the surfaces satisfy in good approximation%
\begin{equation}
\Omega ^{nr,\text{REV}}=\hat{S}^{n}\Omega ^{rp,\text{REV}}\text{ \ and \ }%
\Omega ^{wr,\text{REV}}=\hat{S}^{w}\Omega ^{rp,\text{REV}}=(1-\hat{S}%
^{n})\Omega ^{rp,\text{REV}}
\end{equation}%
The pressure of the REV from Eq. \ref{eq:PVREV} can then be written as: 
\begin{equation}
p=[p^{n}\hat{S}^{n}\phi +p^{w}(1-\hat{S}^{n})\phi +p^{r}(1-\phi )]-[\hat{S}%
^{n}\gamma ^{nr}+(1-\hat{S}^{n})\gamma ^{wr}]\frac{\Omega ^{rp,\text{REV}}}{%
V^{\text{REV}}}-\gamma ^{nw}\frac{\Omega ^{nw,\text{REV}}}{V^{\text{REV}}}
\label{G.1}
\end{equation}%
Contact-line contributions were again not taken along, for simplicity. A
consequence of the porous medium being homogeneous is that $\Omega ^{rp,%
\text{REV}}/V^{\text{REV}}$\ is the same everywhere.
The ratio can be used as a measure of the average curvature of the pore
surface, as will be explained below.

The volume-averaged contributions to the pressure from the homogeneous phases is given the symbol $\bar{p}$: 
\begin{eqnarray}
\bar{p} &=&p^{n}\hat{S}^{n}\phi +p^{w}(1-\hat{S}^{n})\phi +p^{r}(1-\phi ) 
\notag \\
&=&\left( p^{n}-p^{w}\right) \hat{S}^{n}\phi +p^{w}\phi +p^{r}(1-\phi )
\label{p_av}
\end{eqnarray}%
The first term in the last equality shows that the saturation gives
an important contribution to the volume-averaged pressure. The  contributions of the 
2nd and 3rd terms are due to $p^{w}$\ and $p^{r}$.  These contributions are usually constant.  

The surface-averaged contributions to the pressure are likewise given a separate symbol: 
\begin{equation}
\bar{p}^{c}=[\hat{S}^{n}\gamma ^{nr}+(1-\hat{S}^{n})\gamma ^{wr}]\frac{%
\Omega ^{rp,\text{REV}}}{V^{\text{REV}}}+\gamma ^{nw}\frac{\Omega ^{nw,\text{%
REV}}}{V^{\text{REV}}}  \label{eq:pc}
\end{equation}%
The contribution of $\bar{p}^{c}$ to the pressure, $p$, is often called the capillary pressure. With an (approximately) cylindrical pore geometry, we can define the average radius of the pores by%
\begin{equation}
\overline{r}\equiv \frac{2V^{p,\text{REV}}}{\Omega ^{rp,\text{REV}}}
\end{equation}%
This may be a good assumption in the absence of film formations. By introducing 
$\overline{r}$ into Eq.\ref{eq:pc} we obtain the capillary pressure
\begin{equation}
\bar{p}^{c}=\left( \gamma ^{nr}-\gamma ^{wr}\right) \hat{S}^{n}\frac{2\phi }{%
\overline{r}}+\gamma ^{wr}\frac{2\phi }{\overline{r}}+\gamma ^{nw}\frac{%
\Omega ^{nw,\text{REV}}}{V^{\text{REV}}}  \label{eq:pc2}
\end{equation}%
Again the first term shows that saturation gives an important
contribution. The 2nd term only depends on the
temperature and is usually constant. The 3rd term is proportional to the
surface area of the fluid-fluid interface. In many experiments this surface
area is much smaller than $\Omega ^{rp,\text{REV}}$. When that is the case,
this term is negligible. 

The effective pressure of the REV is thus, for short: 
\begin{equation}
p=\bar{p}-\bar{p}^{c}  \label{eq:Wash}
\end{equation}%
The three equations above give an expression for the REV pressure $p$ 
for the example system. 

Approximations should be tested with the more detailed expressions. 
To estimate the size of the various contributions, it is convenient to use
mechanical equilibrium for the contact line and for the surface, although this condition may not apply to the REV, not even under stationary flow conditions. 
With a balance of forces at the three-phase contact lines, Young's law applies for the
surface tensions: $\gamma ^{nr}-\gamma ^{wr}=\gamma ^{nw}$ $\cos \theta $,
where $\theta $ is the (average) contact angle. When there is furthermore
mechanical equilibrium at the fluid-fluid interfaces, the pressure difference between the fluid is given by Young-Laplace's law, $p^{n}-p^{w}=2\gamma ^{nw}\cos \theta /\bar{r}$.  

In the single-fluid ($w$) case, Eq.\ref{G.1} simplifies. The volume-averaged contribution becomes, 
\begin{equation}
\bar{p}=p^{w}\phi +p^{r}(1-\phi )  \label{A.4c}
\end{equation}
and by introducing Eq.\ref{eq:pc2}, we obtain 
\begin{equation}
p=p^{w}\phi +p^{r}(1-\phi )-\gamma ^{wr}\frac{2}{\bar{r}}\phi   \label{A.4b}
\end{equation}

In this section, we have defined in detail what we mean by the pressure of a REV. We have found, using the grand potential, that it can be regarded as result of volume- and surface average properties. These contributions enter the driving force in Eqs. \ref{E.7c} and \ref{D.1}, to be further discussed below.  

\subsection{The pressure difference as driving force across a porous medium}

The driving force for volume flow is the gradient of the REV pressure. 
To measure the pressure $p$ inside the REV is difficult. The
pressure in the fluid phases adjacent to the porous medium can be determined. 
Tallakstad et al. \cite{Tallakstad2009} defined
the measured pressure difference, $\Delta p^{\prime}$, at steady state, 
as an average over the value $\Delta p(t)$ over the time of measurement: 
\begin{equation}
\Delta p^{\prime }=\frac{1}{t_{\mathrm{e}}-t_{\mathrm{s}}}\int_{t_{\mathrm{s}%
}}^{t_{\mathrm{e}}}\Delta p(t)dt
\end{equation}%
Here $t$ is the time and $\Delta$ refers to the extension of the system.
Subscript 's' denotes the start and 'e' denotes the end of the measurement.
We will take this pressure difference as our $\Delta p$. 

The pressure differences $\Delta p^{w}$ and $\Delta p^{n}$ can also be
found when there is continuity in the fluids, $w$ and $n$, respectively.

By taking the difference between the inlet and outlet in Eq.\ref{eq:Wash}, we have an interpretation of the pressure difference;
\begin{equation}
\Delta p=\Delta \bar{p}-\Delta \bar{p}^{c}  \label{eq:Wash2}
\end{equation}%
The question is now how we can assess the right-hand side of this equation, using Eqs. \ref{p_av} and \ref{eq:pc}.   

\subsubsection{Large pressure differences}

When the pressure drop across the porous plug is large compared to the
capillary pressure contribution, the surface contributions and therefore $%
\bar{p}^{c}$ can be neglected. Furthermore $p^{n}=p^{w}$. In the pressure
difference, the terms with constant $\phi $ and $p^{r}$ disappears, and
the pressure difference is: 
\begin{equation}
\Delta p=\Delta {\bar{p}}=\phi \Delta {p^{w}}  \label{eq:lpd}
\end{equation}%
The pressure is applied to the whole cross-sectional area. This explains that 
the net driving force becomes a fraction, $\phi$, of $\Delta p^{w}$. In other words, the force applies to the fraction $\phi $ of the pore area.

The conditions leading to Eq.\ref{eq:lpd} are common in the laboratory. Some
numerical values for the air-glycerol system, \cite{Erpelding2013}, can
illustrate when the conditions apply. The value of $2\phi \gamma ^{wr}/\bar{r}$ is
of the same order of magnitude as $p^{c}\hat{S}^{n}\phi $ (400 Pa) when the
surface tension $\gamma =\ $6.4 $\cdot $ 10$^{-2}$N m$^{-2}$, the average
pore radius $\bar{r} $ = 0.2 mm and the porosity $\phi $ = 0.63. A typical
value of $\Delta \bar{p}$ in the experiments is close to 30 kPa, which is
far from the limit where capillary effects are significant.

\subsubsection{Small pressure differences}

For small capillary numbers the effective pressure drop across a porous plug
is comparable to or smaller than the capillary pressure. Surface
contributions need be taken into account. Equation \ref{eq:Wash2} gives the
effective pressure difference. When we can assume a constant average 
radius $\bar{r}$, and constant porosity, we obtain 
\begin{equation}
\Delta p=\Delta \bar{p}-\frac{2\phi }{\bar{r}}\Delta \left[ \left( \gamma
^{nr}-\gamma ^{wr}\right) \hat{S}^{n}+\gamma ^{wr}\right] -\Delta \left(
\gamma ^{nw}\frac{\Omega ^{nw,\text{REV}}}{V^{\text{REV}}}\right) 
\end{equation}%
A fluid will be transported when the surface tensions of the fluids with the
wall are different and there is a difference in the saturation. 
When there is only one fluid in the porous medium, cf. Eq.\ref{A.4b}, and we have constant $\bar{r}$ and porosity,  the pressure difference becomes 
\begin{equation}
\Delta p=\Delta \bar{p}-\frac{2\phi }{\bar{r}}\Delta \gamma ^{wr}
\end{equation}
The last term can lead to mass transport, when the surface tension changes.

\section{Constitutive equations}

The constitutive equations follow from the entropy production. We present these on differential form 
for two incompressible flows. From Eq. \ref{E.7c} we have:
\begin{eqnarray}
J_{q}^{\prime } &=&l_{qq}\frac{\partial }{\partial x}(\frac{1}{T})-l_{qw}%
\frac{1}{T}\frac{\partial \mu _{w,T}}{\partial x}-l_{qn}\frac{1}{T}\frac{%
\partial \mu _{n,T}}{\partial x}  \notag \\
J_{w} &=&l_{wq}\frac{\partial }{\partial x}(\frac{1}{T})-l_{ww}\frac{1}{T}%
\frac{\partial \mu _{w,T}}{\partial x}-l_{wn}\frac{1}{T}\frac{\partial \mu
_{n,T}}{\partial x}  \notag \\
J_{n} &=&l_{nq}\frac{\partial }{\partial x}(\frac{1}{T})-l_{nw}\frac{1}{T}%
\frac{\partial \mu _{w,T}}{\partial x}-l_{nn}\frac{1}{T}\frac{\partial \mu
_{n,T}}{\partial x}  \label{E1}
\end{eqnarray}
We can also use Eq. \ref{D.1}\ and obtain
\begin{eqnarray}
J_{q}^{\prime } &=&l_{qq}\frac{\partial }{\partial x}\left( \frac{1}{T}%
\right) -l_{qp}\frac{1}{T}\frac{\partial }{\partial x}p-l_{qd}\frac{\rho _{w}%
}{T}\frac{\partial \mu _{w}^{c}}{\partial x}  \notag \\
J_{V} &=&l_{pq}\frac{\partial }{\partial x}\left( \frac{1}{T}\right) -l_{pp}%
\frac{1}{T}\frac{\partial }{\partial x}p-l_{pd}\frac{\rho _{w}}{T}\frac{%
\partial \mu _{w}^{c}}{\partial x}  \notag \\
J_{D} &=&l_{dq}\frac{\partial }{\partial x}\left( \frac{1}{T}\right) -l_{dp}%
\frac{1}{T}\frac{\partial }{\partial x}p-l_{dd}\frac{\rho _{w}}{T}\frac{%
\partial \mu _{w}^{c}}{\partial x}  \label{D1}
\end{eqnarray}
The flux-force relations are linear on this level. The two conductivity matrices can be expressed in each other. The element $l_{qq}$ is the same in both formulations. When the REV can be regarded as a thermodynamic state 
\cite{Erpelding2013,Savani2017}, we can expect that the conductivity matrices on this local form are symmetric, or that the Onsager relations apply, $l_{ij}=l_{ji}$.  Flekkøy \emph{et al.} \cite{Flekkoy2017} and Pride \emph{et al.}\cite{Pride2017} argued that the Onsager relations are obeyed also on the REV level. 
Experimental proof for the Onsager relations of two-phase flow in porous media does not yet exist, 
however, and we shall see below how this possibly can be achieved.

We have discussed above how the overall driving forces can be determined.  
We need to integrate across the system, in order to study their effect on experiments. 
We integrate the linear laws \ref{D1} across the REV, and obtain
\begin{eqnarray}
J_{q}^{\prime } &=&L_{qq}\Delta (\frac{1}{T})-L_{qw}\frac{1}{T}\Delta \mu
_{w,T}-L_{qn}\frac{1}{T}\Delta \mu _{n,T}  \notag \\
J_{w} &=&L_{wq}\Delta (\frac{1}{T})-L_{ww}\frac{1}{T}\Delta \mu _{w,T}-L_{wn}%
\frac{1}{T}\Delta \mu _{n,T}  \notag \\
J_{n} &=&L_{nq}\Delta (\frac{1}{T})-L_{nw}\frac{1}{T}\Delta \mu _{w,T}-L_{nn}%
\frac{1}{T}\Delta \mu _{n,T}  \label{E2}
\end{eqnarray}%
and%
\begin{eqnarray}
J_{q}^{\prime } &=&L_{qq}\Delta \left( \frac{1}{T}\right) -L_{qp}\frac{1}{T}%
\Delta p-L_{qd}\frac{\rho _{w}}{T}\Delta \mu _{w}^{c}  \notag \\
J_{V} &=&L_{pq}\Delta \left( \frac{1}{T}\right) -L_{pp}\frac{1}{T}\Delta
p-L_{pd}\frac{\rho _{w}}{T}\Delta \mu _{w}^{c}  \notag \\
J_{D} &=&L_{dq}\Delta \left( \frac{1}{T}\right) -L_{dp}\frac{1}{T}\Delta
p-L_{dd}\frac{\rho _{w}}{T}\Delta \mu _{w}^{c}  \label{D2}
\end{eqnarray}%
Here $L_{ij}\equiv l_{ij}/l$ and  $l$\ is the length of the REV, and the driving forces are defined by Eqs. 
\ref{eq:Wash2} and \ref{eq:mu_c}. 

The coefficients may become dependent on the force through the integration. To illustrate this, consider an example. Two fluids in a capillary were studied, and linear laws were first written for each of them on the pore-level \cite{Sinha2013}.  The average velocity of a bubble was found by integration from the pore- to the macro-level, using the configurational distribution $\Pi (x_{b})$, of the position, $x_b$, of the fluid interfaces. The capillary pressure depended on $x_{b}$. The averaging procedure gave the conductivity as a function of ($\Delta \bar{p}$ - $\Delta p^c$) in the terminology of this paper. 

In the remainder of the paper we will discuss experimental conditions that allow us to determine these coefficients.  The presentation follows closely the derivation of Stavermann \cite{Stavermann1952} and Katchalsky and coworkers \cite{Katchalsky1965} for
transport in discrete systems of polymer membranes, see also F{/o}rland \cite{Forland1988}. We refer to these works for further definitions of transport coefficients. 

\subsection{Isothermal, single fluid flow.}

For an isothermal single fluid $w,$ flowing inside a porous medium, the
entropy production \ref{D2} has one term; the volume flow times the negative
pressure difference over the temperature. By including the constant
temperature in the transport coefficient, we obtain the common linear law. With the
permeability $L_{p}$, we write 
\begin{equation}
J_{V}=-L_{p}\Delta p  \label{H.1}
\end{equation}%
where $L_{p}\equiv L_{pp}/T$. The permeability is normally a function of state
variables (pressure, temperature). In the hydrodynamic regime it is a
function of viscosity, $L_{p}=L_{p}(p,T,\eta )$.  
By introducing the new expression for the pressure \ref{A.4b}, we
obtain 
\begin{equation}
J_{V}=-L_{VV}\left( \Delta p^{w}-\frac{2}{\bar{r}}\Delta \gamma ^{wr}\right) 
\label{K.1}
\end{equation}%
When the permeability and porosity are constant, $L_{VV}\equiv L_{p}\phi $. The
equation predicts a threshold value for flow if there is a (significant)
change in the surface tension across the REV. Transport will take place,
when $\Delta p^{w}>{2\Delta \gamma ^{wr}}/{\bar{r}}$. The permeability $%
L_{VV}$ is inversely proportional to the viscosity $\eta $ of the fluid in the hydrodynamic regime.
Interestingly, Boersma et al. \cite{Boersma1973} and Miller et al. \cite%
{Miller1963} plotted the volume flow versus the hydrostatic pressure
difference $\Delta p^{w}$ and found a deviation from Darcy's law in the form
of a pressure threshold, for water or water solutions in clay. They offered
no explanation for this. Also Bernadiner et al. \cite{Bernadiner1994} and
Swartzendruber \cite{Swartzendruber1962} plotted the volume flow of water
solution $J_{V}$ versus the pressure gradient in sandstone with low clay
content \cite{Bernadiner1994}, and in NaCl-saturated Utah bentonite \cite%
{Swartzendruber1962}. The thresholds that they observed depended on the content 
of salt in the permeating solution. They explained the thresholds by water
adsorption and pore clogging by colloids \cite{Bernadiner1994}. According to
Eq.\ref{K.1}, a varying surface tension (due to a varying adsorption and
clogging) might explain the existence of a threshold or a non-linear flux-force relation. The dependence of the coefficient $L_{VV}$ on the threshold pressure can also have other explanations, cf. the case described above \cite{Sinha2013}. 
This non-linearity does not prevent the use of non-equilibrium thermodynamics. 

\subsection{Isothermal flow of two components}

The entropy production in Eq.\ref{D.1} has two terms when two immiscible
components flow at isothermal conditions. We choose the formulation that has
variables $J_{V}$ and $J_{D}$; volume flux and interdiffusion flux,
respectively. Equation \ref{D2} gives then: 
\begin{eqnarray}
J_{V} &=&-L_{pp} \Delta p-L_{pd} \left( \rho _{w}\Delta
\mu _{w}^{c}\right)   \notag \\
J_{D} &=&-L_{dp} \Delta p-L_{dd} \left( \rho _{w}\Delta
\mu _{w}^{c}\right)   \label{C.2}
\end{eqnarray}%
where $L_{ij}^{\prime }\equiv L_{ij}/T$. The coefficients reflect, as above,
the mechanism of flow (pressure, diffusion). Four experiments can be done to determine the four coefficients. There are only three independent coefficients. When four experiments are done, we can check the Onsager relations.

\subsubsection{The hydraulic permeability}

The (hydraulic) permeability is a main coefficient. 
By introducing the driving force for the volume flow from \ref{eq:Wash2}, we obtain
\begin{equation}
J_{V}=-L_{pp}\Delta {p}=-L_{pp}\Delta (\bar{p}-\bar{p}^{c})
\label{eq:5.1}
\end{equation}
With the present definition of variables the equation applies to the overall behavior of the system. A plot of $J_{V}$ vs. $\Delta \bar{p}$ may show a threshold. 
This threshold has more contributions
than in the single component system, as there are contributions to the
pressure from surface and line energies. A threshold may be detectable at low capillary numbers. 

The hydraulic or volume permeability, 
$L_{pp}$, is found by measuring the volume flow caused by the overall
pressure difference at uniform composition; 
\begin{equation}
L_{pp}=-\left( \frac{J_{V}}{\Delta p}\right) _{d\mu _{w}^{c}=0}
\end{equation}
The coefficient is a function of the saturation 
$L_{pp}=L_{pp}(p,T,\eta ,\hat{S_{w}})$. 
In the hydrodynamic regime, the coefficient can be
modeled, assuming Poiseuille flow and the effective
viscosity $\eta ^{\mathrm{eff}}=\eta _{w}\hat{S}_{w}+\eta _{n}\hat{S}_{n}$ 
\cite{Sinha2013}.

\subsubsection{The interdiffusion coefficient}

The main coefficient $L_{dd}$ is an interdiffusion coefficient. It is
defined at uniform pressure from the difference flux created by a difference
in saturation; 
\begin{equation}
L_{dd}=-\left( \frac{J_{D}}{\rho _{w}\Delta \mu _{w}^{c}}\right) _{\Delta
p=0}=- \frac{W_{w}}{\phi RT \rho_{w}^0} \left( \frac{J_{D}}{\Delta \hat{S}%
^{w}}\right) _{\Delta p=0}
\end{equation}%
where we used Eq.\ref{eq:mu_c} for the driving force.

\subsubsection{The coupling coefficients} 

The coupling coefficients in Eqs.\ref{C.2} express that a separation of components can be caused by a pressure gradient ($L_{dp}$) and that a volume flow can be promoted by a gradient in saturation ($L_{pd}$). 

Consider first the determination of $L_{dp}$. 
A pressure gradient may build as a consequence of a
difference in composition  \cite{Forland1988}. 
The volume flux continues until a balance of forces is reached: 
\begin{equation}
\Delta p = -\frac{L_{pd}}{L_{pp}}\rho _{w}\Delta {\mu _{w}^{c}} 
\end{equation}%
From the force-balance across the system, we obtain: 
\begin{equation}
\left( \frac{\Delta p}{\Delta \hat{S}^{w}}\right) _{J_{V}=0}=-\frac{L_{pd}}{%
L_{pp}} \phi \frac{\rho_{w}^0 RT}{W_{w}}
\end{equation}%
This condition can be used to find the unknown coupling coefficient, once
the hydraulic permeability is known. 

The remaining coupling coefficient can be found from the flux ratio, $r$, 
that has been called the reflection coefficient $r$, 
see F{/o}rland \cite{Forland1988}. At constant saturation, we have 
\begin{equation}
r=-\left( \frac{J_{D}}{J_{V}}\right) _{\Delta \mu _{w}^{c}=0}=-\frac{L_{dp}}{L_{pp}}
\end{equation}
We are now in a position to compare $L_{pd}$ and $L_{dp}$. The state of the system must be (approximately) the same, when the comparison is made. 

\subsection{Non-isothermal flow of two components}

The set of equations \ref{D2} describe non-isothermal flow in porous media.
The coefficients, $L_{pp}$, $L_{pd}=L_{dp}$, $L_{dd}$ in the lower
right-hand side corner of the conductivity matrix, were discussed above. The
new coefficients are those related to heat transport. The coefficient $L_{qq}
$ represent the Fourier type heat conductivity at uniform composition and
pressure. The coefficients $L_{pq}$ and $L_{dq}$ are coupling coefficients.

Non-zero coefficients $L_{pq}$ and $L_{dq}$ mean that we can obtain
separation in a temperature gradient. Injection of cold water into warm reservoirs
may thus lead to separation. Likewise, a pressure difference can arise from
a temperature difference. This is called thermal osmosis \cite{Barragan2017}. 

Separation caused by a thermal driving force was observed in clay-containing soils
where water was transported in clay capillaries against a pressure gradient.
The coefficient, measured at constant pressure, was called the segregation
potential \cite{Konrad1999}. The coefficient can be obtained from Eq. \ref{D2}, setting $\Delta p=0$ and $\Delta \mu _{w}^{c}=0$  ($\Delta \hat{S}^w$) in the second
line. We obtain 
\begin{equation}
\left( \frac{J_{V}}{\Delta T}\right) _{\Delta p=0,\Delta \mu _{w,T}=0}=-%
\frac{1}{T^{2}}L_{pq}
\end{equation}
This coefficient can also be found from steady state conditions, when the 
thermal gradient is balanced by a gradient in saturation (chemical potential) 
\begin{eqnarray}
L_{pq}\frac{1}{T^{2}}\Delta T &=&-L_{pd}\rho _{w}\Delta \mu _{w}^{c} =
-L_{pd} \phi \frac{RT\rho _{w}^0}{W_{w}} \Delta \hat{S}^{w}  \notag \\
\left( \frac{\Delta \hat{S}^{w}}{\Delta T}\right) _{J_{D}=0} &=&-\frac{W_{w}}{RT^{3}\rho _{w}^0}\frac{L_{pq}}{L_{pd}}
\end{eqnarray}
Determination of $L_{pq}$ requires knowledge of $L_{pd}$. 

The coupling coefficient $L_{qp}$ can be found by measuring the heat flux that accompanies the volume flux for constant composition and at isothermal conditions.  
\begin{equation}
\left( \frac{J_q'}{J_V} \right)_{\Delta T = 0, \Delta \mu_w^c=0 } = \frac{L_{qp}}{L_{pp}}
\end{equation}
These effects have, to the best of our knowledge, so far not been measured for porous media. For homogeneous media they are well known. 

\section{Discussion and Conclusion}


We have used the new formulation of the entropy production \cite{Kjelstrup2018} to find constitutive equations for flow of two immiscible 
fluids in a porous medium under uniform or varying temperature, pressure and
composition. Several of the equations are new in the context of porous
media, but they follow well documented tracks in classical non-equilibrium thermodynamics \cite{Stavermann1952,Katchalsky1965,Forland1988,Kjelstrup2018}. Experimental observations exist on single fluid flow, that give support to the theoretical description. 

We have defined the pressure of the REV, and used the definition to define
the pressure part of the driving force.  The force obtains contributions from the surface -and, in principle also line - tensions of the system. 
This distinguishes the present formulation from their counterpart for homogeneous systems \cite{Stavermann1952,Katchalsky1965,Forland1988,Kjelstrup2018}. 
We have seen that
surface contributions can be spelled out for varying conditions, under the
assumptions that the additive properties of the REV are Euler homogeneous of the
first order. Doing this, we have been able to explain for instance
deviations from Darcy's law, the occurrence of threshold pressures. We have
defined the transport equations on the macroscopic scale, and pointed at
possibilities to describe non-isothermal phenomena. 
As for instance Eq.\ref{G.1} shows, there is a multitude of scenarios that can be further investigated, and used to check the theory, cf. subsections 5.1-5.3. The expressions open up the possibility to test the thermodynamic models in use, for their compatibility with the second law.


The basic assumption that the REV set of basis variables are Euler homogeneous functions of degree one means in its essence that \emph{one} temperature, 
\emph{one} pressure
and \emph{one} chemical potential per component can be defined for the REV. The
properties can vary in time and space, and can therefore also be used for a transient description.  We did not consider surface areas or
their curvature as independent variables, but these may be included as variables. 
In that sense, our description be equivalent to a description using Minkovsky functionals \cite{McClure2018}. The state of the REV is a
thermodynamic state, meaning that the REV is ergodic. Some evidence already supports
this idea \cite{Erpelding2013}, \cite{Savani2017}, originally proposed by Hansen and Ramstad \cite{Hansen2009} and Tallakstad and coworkers \cite{Tallakstad2009}.

We have often used a specific case to illustrate relations; the non-isothermal flow of
one or two immiscible fluids in a non-deformable medium. It is straight
forward to include more terms in the chemical potential (e.g. gravity). To
include stress fields or other fields is more problematic.

Flow of two isothermal, immiscible fluids in a porous medium has often been
described by Darcy's law, using the relative permeability concept. The
seepage velocities $v_{n}$ and $v_{w}$ are related to fluxes used here by $%
v_{n}=J_{n}V_{n}$ and $v_{w}=J_{w}V_{w}$. The expressions for the seepage velocities must be contained or be equivalent to the expressions given here, using the condition of invariance for the entropy production. A comparison can elucidate assumptions that are made.  Hilfer and Standnes et al. \cite{Hilfer1998,Standnes2017} gave a set of linear relations for the seepage velocities. Their driving forces were the gradients in the single component pressures, obtained by pressure measurements in the single phases. Their description implies \emph{e.g} that the composition is uniform. 

We have seen through these examples how non-equilibrium thermodynamic theory
gives a fundamental basis to the constitutive equations. By demanding that
transport properties obey entropy production invariance and Onsager
symmetry, we can also find relations between variables used so far.

\subsection*{Acknowledgment}

The authors are grateful to the Research Council of Norway through its
Centres of Excellence funding scheme, project number 262644, PoreLab. Per
Arne Slotte is thanked for stimulating discussions.


{}

\section*{Symbol lists}


\text{\textbf{Table 1. Mathematical symbols, superscripts, subscripts}} 
\newline
\begin{tabular}{ll}
{Symbol} & {Explanation} \\ \hline
$c$ & superscript meaning capillary pressure \\ 
$d$ & differential \\ 
$\partial$ & partial derivative \\ 
$\Delta$ & change in a quantity or variable \\ 
$\Delta_{f,t}$ & the change is taken from $f$ on the right \\ 
& to $t$ on the left hand side \\ 
$\Sigma$ & sum \\ 
$i$ & subscript meaning component i \\ 
$m$ & number of fluids \\ 
$n$ & subscript meaning non-wetting fluid \\ 
$w$ & subscript meaning wetting fluid \\ 
$p$ & superscript meaning pore \\ 
REV & abbreviation meaning representative elementary volume \\ 
$r$ & superscript meaning rock, solid matrix of medium \\ 
s & superscript meaning interface \\ 
$u$ & subscript meaning internal energy \\ 
$\alpha,\beta$ & superscripts meaning surface between phases $\alpha$ and $%
\beta$ \\ 
$\alpha,\beta,\delta $ & superscripts meaning contact line between phases $%
\alpha,\beta,\delta $ \\ 
$\theta$ & contact angle, average \\ \hline
\end{tabular}

\newpage

\text{\textbf{Table 1. Latin symbols}} \newline
\begin{tabular}{lll}
{Symbol} & {Dimension} & Explanation \\ \hline
$G$ & J & Gibbs energy \\ 
$M$ & kg & mass \\ 
$d$ & m & pore length \\ 
$H_{i}$ & J.kg$^{-1}$ & partial specific enthalpy of i \\ 
$J$ & kg.s$^{-1}$.m$^{-2}$ & mass flux \\ 
$J_{q}^{\prime }$ & J.s$^{-1}$.m$^{-2}$ & sensible heat flux \\ 
$l$ & m & characteristic length of representative elementary volume \\ 
$L$ & m & characteristic length of experimental system \\ 
$L_{ij}$ &  & Onsager conductivity \\ 
$p$ & Pa & pressure of REV \\ 
$Q$ & m$^{3}$.s$^{-1}$ & volume flow \\ 
$\bar{r}$ & m & average pore radius \\ 
$S$ & J.K$^{-1}$ & entropy \\ 
$s$ & J.K$^{-1}$.m$^{-3}$ & entropy density \\ 
$S_{i}$ & J kg$^{-1}$.K$^{-1}$ & partial specific entropy of $i$ \\ 
$\hat{S_{i}}$ &  & degree of saturation of $i$, $\equiv V_{\mathrm{i}}/V$ \\ 
$T$ & K & temperature \\ 
$t$ & s & time \\ 
$U$ & J & internal energy \\ 
$u$ & J.m$^{-3}$ & internal energy density \\ 
$V$ & m$^{3}$ & volume \\ 
$V_{i}$ & m$^{3}$.kg$^{-1}$ & partial specific volume \\ 
$x$ & m & axis of transport \\ 
$W_{i}$ & kg.mol$^{-1}$ & molar mass of $i$%
\end{tabular}

\newpage

\text{\textbf{Table 2. Greek symbols, continued}} \newline
\begin{tabular}{lll}
{Symbol} & {Dimension} & Explanation \\ \hline
$\alpha$ &  & superscripts meaning a phase \\ 
$\beta$ &  & superscript meaning an interface \\ 
$\delta$ &  & superscript meaning a contact line \\ 
$\phi$ &  & porosity of porous medium \\ 
$\gamma$ & N.m$^{-1}$ & surface tension  \\ 
$\mu_{\mathrm{i}}$ & J.kg$^{-1}$ & chemical potential of i \\ 
$\rho_{\mathrm{i}} $ & kg.m$^{-3}$ & density, $\equiv M_{\mathrm{i}}/V_i $
\\ 
$\sigma$ & J.s$^{-1}$.K$^{-1}$.m$^{-3}$ & entropy production in a
homogeneous phase \\ 
$\Omega$ & m$^2$ & surface or interface area \\ 
&  & 
\end{tabular}
\end{document}